# minimUML: A Minimalist Approach to UML Diagraming for Early Computer Science Education

SCOTT A. TURNER, MANUEL A. PÉREZ-QUIÑONES AND STEPHEN H. EDWARDS
Virginia Polytechnic Institute and State University

The Unified Modeling Language (UML) is commonly used in introductory Computer Science to teach basic object-oriented design. However, there appears to be a lack of suitable software to support this task. Many of the available programs that support UML focus on developing code and not on enhancing learning. Those that were designed for educational use sometimes have poor interfaces or are missing common and important features, such as multiple selection and undo/redo. There is a need for software that is tailored to an instructional environment and has all the useful and needed functionality for that specific task.
This is the purpose of minimUML. minimUML provides a minimum amount of UML, just what is commonly used in beginning programming classes, while providing a simple, usable interface. In particular, minimUML was designed to support abstract design while supplying features for exploratory learning and error avoidance. Some of the functionality supported includes multiple selection, undo/redo, flexible printing, cut and paste, and drag and drop. In addition, it allows for the annotation of diagrams, through text or freeform drawings, so students can receive feedback on their work. minimUML was developed with the goal of supporting ease of use, supporting novice students, and a requirement of no prior-training for its use. This paper presents the rationale behind the minimUML design, a description of the tool, and the results of usability evaluations and student feedback regarding the use of the tool.

## 1. INTRODUCTION

The Unified Modeling Language (UML) is commonly used in introductory Computer Science courses as a way to introduce students to object-oriented design. Professors present UML as a way of talking about design problems and as a way to organize problem solutions. In turn, the students create a few diagrams to show their high-level concept of a system. In this paper we present the design, evaluation, and experiences using a minimalist tool to support the use of UML early in the CS curriculum. We call this tool minimUML.

Authors' addresses: Dr. Manuel A. Pérez-Quiñones, Department of Computer Science (0106), Virginia Polytechnic Institute and State University, Blacksburg, VA 24061.

## 1.1 Motivation

In choosing a tool to support UML activities early in the curriculum, we must ensure that it meets the educational needs of students at this level. It must provide at least a subset of UML that meets the needs of the typical types of design problems given in introductory Computer Science classes. It must provide for abstract and iterative design. It must support exploration during the design process and it must provide an interface that is simple enough to be used during a class or a lab with only a brief introduction. Our goal is to evaluate existing systems against these requirements and to create a tool that will adequately meet these needs.

Surprisingly, there is an apparent lack of suitable software to support this task. While there are numerous commercial and free systems that have UML diagraming capabilities, they are, in general, too complex to meet the educational needs of students at this level (Buck and Stucki 2000). Many are focused on creating well-defined diagrams and on code generation. There is an assumption that the user is already familiar with good design principles.

Typically, a very small subset of UML is used in the introductory classes. Many textbooks and professors focus on class diagrams that contain very little detail but show the relationships between the objects through aggregation, association, and inheritance (e.g., Fowler and Scott 1997; Lee and Tepfenhart 2001; Barnes and Kolling 2003; QuickUML 2003; Riley 2003; Horstmann 2004; Lewis and Chase 2004). Since this is the goal of the instruction, there is little need for the tools to support the full UML language. In fact, Alphonce and Ventura (Alphonce and Ventura 2002; Crahen, Alphonce et al. 2002; Alphonce and Ventura 2003) suggest that limiting the amount of UML taught in this situation is helpful in learning design.

Ease of use becomes an important goal if a UML diagraming tool is to be useful early in the Computer Science curriculum. Features for exploratory learning and error avoidance are needed so that the student can try out different ideas without being penalized for going down tangents or making mistakes. To help achieve this goal, the tool should at least support full undo/redo, so students can feel free to experiment with alternative ideas. By keeping the interface simple, students are allowed to learn how to design and not how to use the tool.

The tool must support transferring UML designs out into another work environment. At the very least, the tool should support flexible output features as students might need to print their work, or export it in a graphical format (e.g., PNG or JPG) so that it can be included in a report or used in a presentation. Students also might want to transfer their work to other programs (e.g. text editors, compilers, etc.), by converting the UML designs into code. Since the most used OO languages in the early CS curriculum are Java and C++, it would be appropriate to provide support for both. In some introductory courses at Virginia Tech, students have turned in diagrams created through ASCII art, boxes and arrows in Word, and images that were created in a paint program or scanned. Considering the complexity of the design solution, the students probably spent more time formatting the diagram than designing the solution.

The tool should provide support for feedback and comments within the tool itself. Instructors or peer-reviewers often provide comments to other students so the tool should have some annotation features (Anderson and Shneiderman 1977; Sullivan 1994; Zeller 2000; Gehringer 2001). Finally, the tool should support abstract design, and not coding, by not requiring students to write Java or C++ in their design diagrams (Alphonce and Ventura 2002). There is no urgency in enforcing proper class names, the specification of return types, etc. Those are details that are worked out later in the design of the system, and tools that enforce these up front are inappropriate for the early stages of the design.

This paper presents minimUML, the rationale behind its design, a description of the tool, and the results of usability evaluations and student's feedback. We first define the requirements of an UML tool designed for early CS education. Then we review five currently available UML diagraming tools for their usefulness in this particular educational setting. Next, we present minimUML, our solution to this problem, and describe its features. We present the results of a usability evaluation to improve the design of minimUML and the result of a student survey about their use of minimUML, their design approaches, and the amount of time they spent on learning to use minimUML. Finally, we conclude the paper with a discussion of future work on the tool and the conclusions for our work.

## 2. REQUIREMENTS

When designing a tool, it is important to consider the characteristics of the people it is targeted towards and the situations it will be used in. In general, first year Computer Science students do not have much experience in object-oriented design. Some are accomplished coders, but may have little knowledge about properly structuring programs. Others have no programming skills at all. And the majority of them have little or no idea what UML is. As part of their courses, they may be asked to solve relatively simple design problems consisting of a handful of classes (Lee and Tepfenhart 2002; Horstmann 2004; Lewis and Chase 2005) by making UML diagrams. A typical problem can be found in Lewis and Chase (2005) on page 149 and shown here in Figure 1 as redrawn with minimUML. To prepare students for these assignments, the professors may spend part of a couple of lectures discussing UML and how to represent their designs and then they are set loose to complete their work as homework or as part of a lab session. Students are faced with the three tasks of quickly learning how to structure a program, how to represent it in UML, and how to operate any tool that they might be required to use for the assignment. That could be quite overwhelming. A good tool for this situation should try to limit the amount of effort that go into the second and third tasks to allow more focus on the first. It should also provide a variety of features that can handle common classroom activities.

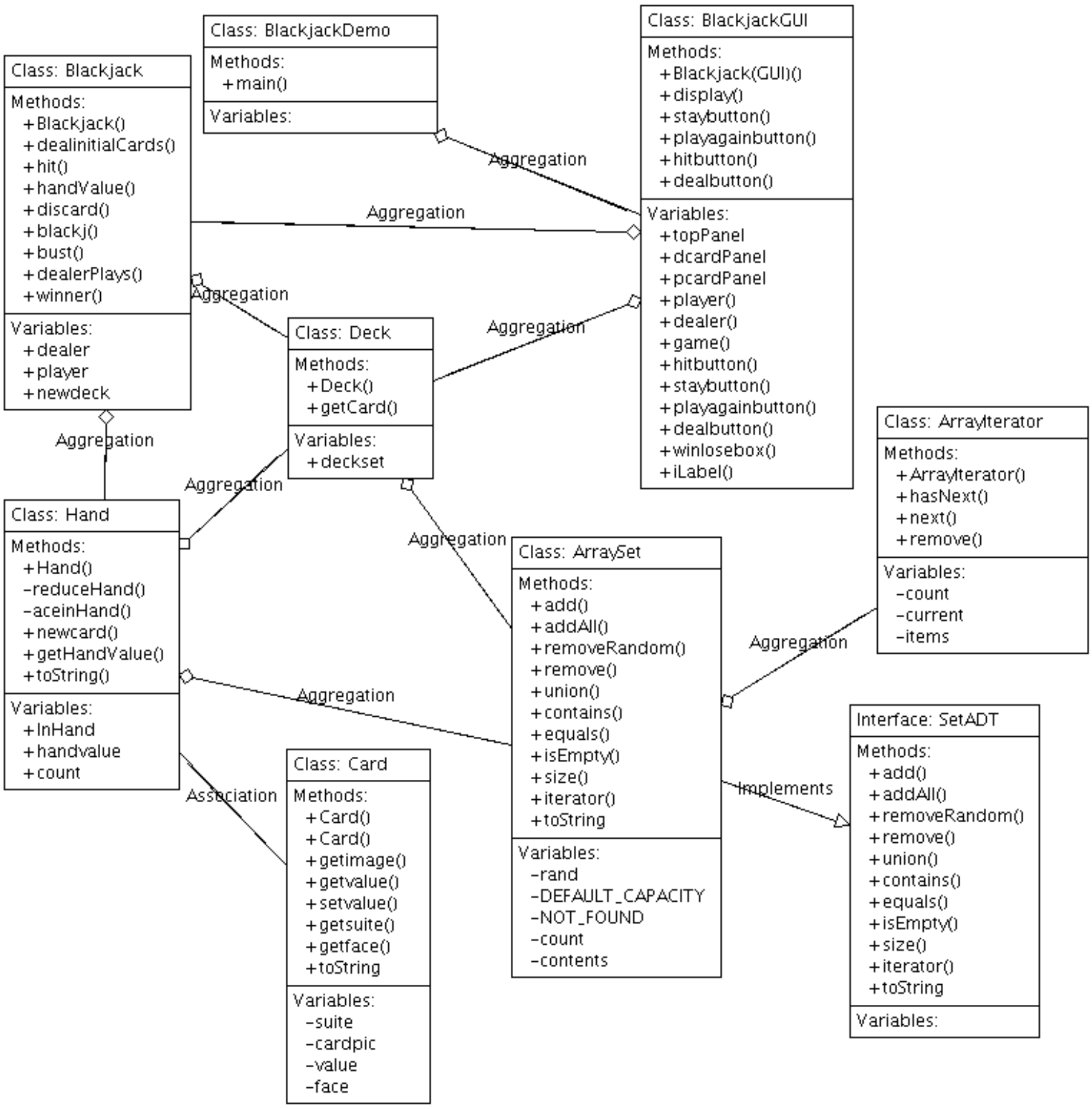

Figure 1: BlackJack Game Design from the Lewis & Chase textbook (2005, 149) as drawn with minimUML

In this setting, the goal is not to teach UML, but to teach the basics of object-oriented design with UML. To do this, **the entire complexity of UML is not needed**. In fact, very little UML is needed to express the solutions to the design problems with which the students will be faced. The solutions typically only involve classes and a few types of relationships between those classes. Nothing else is needed. In this, we are making an argument similar to the one used by QuickUML.

It is also important that the tool should be **useable in a classroom with little or no previous instruction**. The interface should be as simple and easy to use as possible and it should **supply the common functionality users expect from programs**, such as cut and paste and especially **undo/redo**. Moreover, the students should to be able **to try out ideas for their designs without penalty** and to be able **to design in their own ways**. There should not be restrictions imposed on the design by the syntax of a particular programming language.

Support for exploration is another needed component. There is **no need to force students to make early commitments in their designs**. For instance, a student may create a class object and later realize that is should have been an interface. There is little reason to make students explicitly choose a class or an interface when they may only know that they need some kind of object that supports some set of methods. The types of connections between classes face a similar issue. Once again, it is not necessary to insist that the students commit to a particular type of connection when it is created only to have to delete it and start again if it was the wrong type. A good tool should allow them to make tentative choices and help them make changes without needlessly wasting their efforts.

The tool should provide a number of other capabilities to further support classroom activities. The diagrams will need to be submitted in some fashion. Good **printing features** and **ways to export the designs** into standard formats are needed. Support for electronic submission would also help here. The diagrams will be reviewed, by the professor or by other students. **Methods for annotating the assignment** should be included to support this. Finally, the creation of these designs could be the first step of a programming project. The ability to export the classes as code should be included as well. Even better would be to let the student modify both the diagram and the generated code concurrently so that they can work at different levels of abstraction throughout the entire project without having to maintain the pieces individually.

All these features together would make a tool suitable for both the target audience and for the situations it would be used in.

## 3. RELATED WORK

In examining diagraming programs for educational work, we found five UML tools, Violet (2003), UMLet (2004), Dia (2003), ArgoUML (2003), and QuickUML (2003). (URLs for the tools are available in the reference section.) We evaluated the tools based on the subset of UML they support, the simplicity of the user interface, the support for exploratory learning, such as undo/redo facilities, the amount of consistency the interface provided, and how the tool allowed for the avoidance of errors.

When evaluating the tools, we compared their subset of UML to the subset found in some of the available Computer Science textbooks that are intended for introductory object-oriented design. Specifically, we examined the case studies in Horstmann's *Object Oriented Design & Patterns* (2004) and Lewis and Chase's *Java Software Structures: Designing & Using Data Structures* (2004) for examples of typical design problems and the UML used to solve those problems. Lewis and Chase provide an overview of UML (pages 13-17) that describes classes and a few types of relationships between classes. The description provides information about marking classes as abstract or as interfaces and about the formatting of variables and methods with their types and parameters. It is interesting to note that although this formatting information is presented, most of it is not commonly used in the book's diagrams and case studies. They show classes that contain only variable and methods names and little other information (e.g., Lewis and Chase (2005) page 149). This, presumably, is to allow the student to get an idea of the high level design without having to deal with all the details. Horstmann provides a similar overview and presents similar diagrams in terms of complexity. This strengthens our claim that it is not necessary to support all the features of UML, or even all of the formatting options for specific parts of UML, in order to have a tool that is adequate for learning.

### 3.1 Comparison

To better understand the strengths and weaknesses of these UML tools, we compared them across a number of dimensions. The summary of our evaluation is found on Table 1 below. As mentioned earlier, we evaluated the simplicity of the interface, and the amount of undo/redo support provided. We also compared the methods for exporting designs into other formats, such as Java or C++ code or as image files. Other aspects evaluated were: the availability of annotation tools and the possible locations of connections (e.g., between objects or anywhere on the screen). Two of the most important features we looked at were the focus of the tools and the amount of UML they provided as these features help us gauge their use in learning design as opposed to just creating designs. In addition, the presence of features for zooming, multiple selection, printing, and for extending the interface were noted. The final column on Table 1 shows the abilities of minimUML, the tool we created to meet the requirements of early Computer Science education. minimUML is discussed in detail in section 4.

Table 1: Tool Comparison

| | Violet | UMLet | Dia | ArgoUML | QuickUML | minimUML |
|---|---|---|---|---|---|---|
| Simplicity of Interface | Simple | Complex | Complex | Complex | Simple | Simple |
| Undo/redo | None | Full | Some | None | None | Full |
| Printing | To postscript | To PDF | Simple | Simple | Flexible | Flexible |
| Save as image | Yes | Yes | Yes | Yes | Yes | Yes |
| Save as code/text | No | No | No | Yes (Java) | Yes (Java/C++) | Yes/Limited (Java/C++) |
| Annotation tools | Yes | Yes | Yes | Yes | Yes | Yes |
| Focus on design | Yes | No | No | No | Yes | Yes |
| UML covered | Most | Most | Most | Most | Limited | Limited |
| Zoom | Yes | No | Yes | Yes | No | Yes |
| Connection locations | Anywhere | Anywhere | Anywhere | Between classes | Between objects (includes notes) | Between classes |
| Multiple selection | No | Yes | Yes | Yes | Yes | Yes |
| Extensibility | No | Yes | Yes | No | No | No |

As can be seen from the table, the tools cover a wide range of functionality. In general, those that were focused on design had simpler interfaces but lacked some of the nicer features, most notably undo/redo, that the more complex programs provided. Conversely, those that were not focused on design and were more general diagraming or productivity tools typically had more features but did less to support the user in creating valid, abstract diagrams by, for instance, allowing connections to be drawn anywhere or by requiring values to be well specified. (It should be noted that improvements have been made on several of these tools since they were evaluated. For instance, QuickUML is now an Eclipse plug-in called Green (2005) with new features including round trip editing.)

Violet had a wonderfully simple interface but did not have faculties for exploratory learning or ways to avoid errors. It did not have undo/redo, so it was harder to recover from mistakes and it permitted actions that are not logical, such as connections that do not connect anything. It was also missing many features, like multiple selection or better print options, that would make it more convenient to use. UMLet allowed for exploratory learning and for expanding and adapting the user interface but suffered from an extremely poor user interface. It required the user to enter specialized codes to format the objects, like setting the direction of an arrow on a connection, and was missing some very basic UI affordances, like scrollbars for the diagraming area. Dia and ArgoUML provided a great amount of functionality but their complexity made them difficult to use. Dia, a more general purpose diagraming tool, allowed the user to control a large number of settings, such as color and line thickness, for each object which made the interface very complex. Since Dia was not focused solely on UML, it did not provide the user with any support for creating valid diagrams. ArgoUML, on the other hand, ensured valid diagrams by requiring users to make Java-specific design decisions early on. It also suffered from a very cluttered interface. QuickUML attempted to meet the requirement but is missing too many features to fill the need. Its supported subset of UML and its focus are right for academic use, but its interface was found lacking. Like Violet, it lacks undo/redo support. It also allowed the user to make mistakes that could not be recovered from. For instance, objects could be moved off of the screen in such a way that they could not be retrieved. The lack of a suitable UML tool, for an educational context, motivates the creation of one that can meet these goals

## 4. MINIMUML

minimUML is a minimalist (Carroll 1997) approach to UML diagraming. Its purpose is to focus the user on creating object-oriented designs and not on learning the tool. Learning how to use the tool comes as the users use it to perform real tasks, in this case, diagraming.

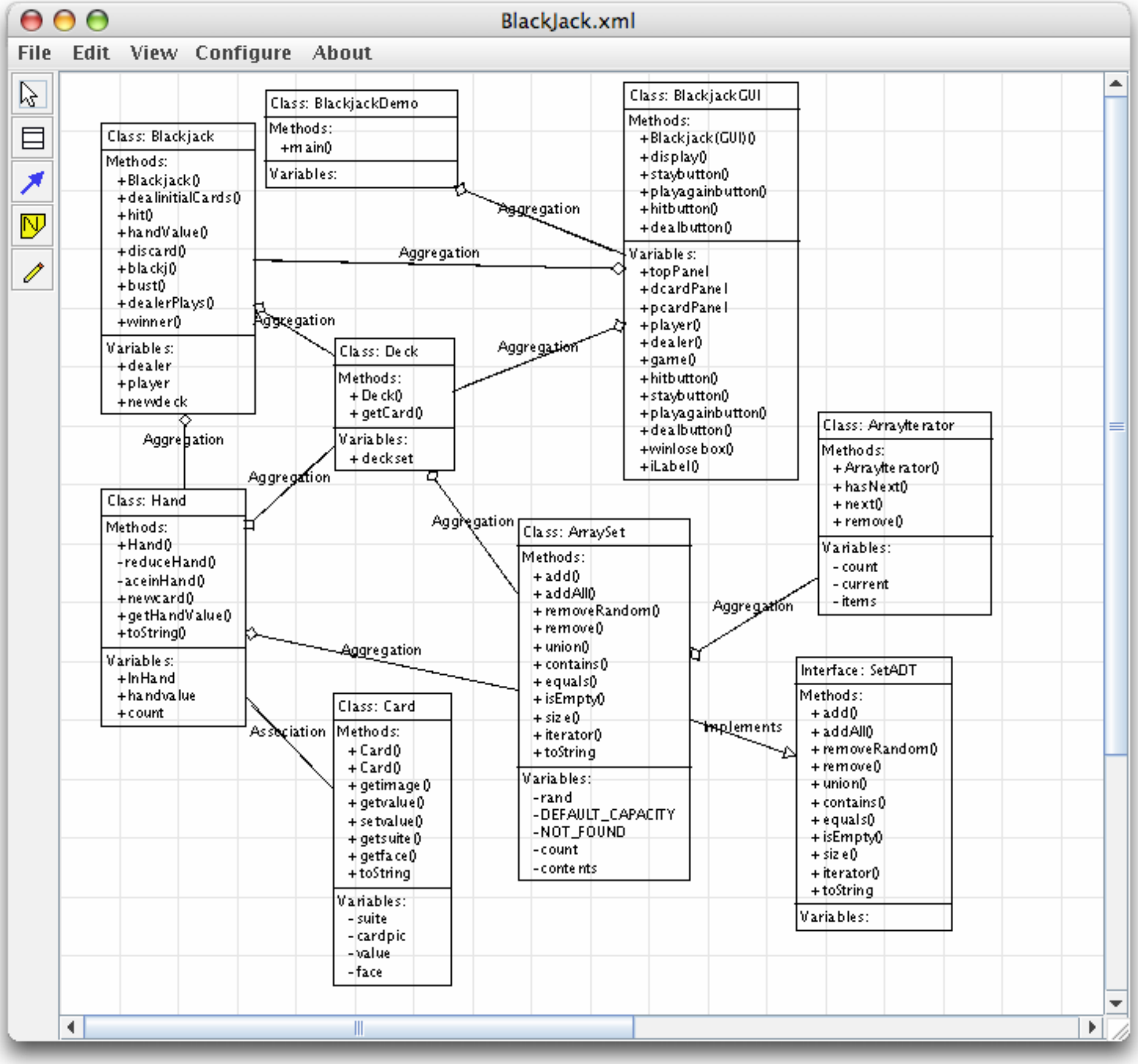

Figure 2: minimUML interface

minimUML provides support for classes, connections between classes, and two types of annotations. For the target audience and for the general types of problems they are solving, as described above, this should be sufficient for their needs. A number of other features, such as undo/redo, cut and paste and drag and drop are supplied by the program to ease the design process. The full interface is shown in Figure 2.

## 4.1 Classes

Classes provide areas for the name of the class, methods, and variables. There are few restrictions imposed on these values other than the class name must be unique to the diagram. While we could have enforced a particular coding style or language syntax, it limits the usefulness of the program to do so. Since both C++ and Java are commonly used to teach OO design, it is not prudent to support one over the other. In addition, this kind of enforcement takes time away from the design process. If the goal is to create a diagram, then stopping the user to add a comma or a semicolon is counterproductive. The compiler will catch these problems later as the design is implemented. This has the fortunate side effect of allowing the user to design the class in iterative steps. Variables and methods can be defined vaguely in the beginning and fleshed out as the design coalesces. There is no need to have the types and number of parameters well defined as items are added to a class. Users are allowed to explore ideas and work in their own ways without many restrictions. Figure 3 shows the selected view of class in minimUML.

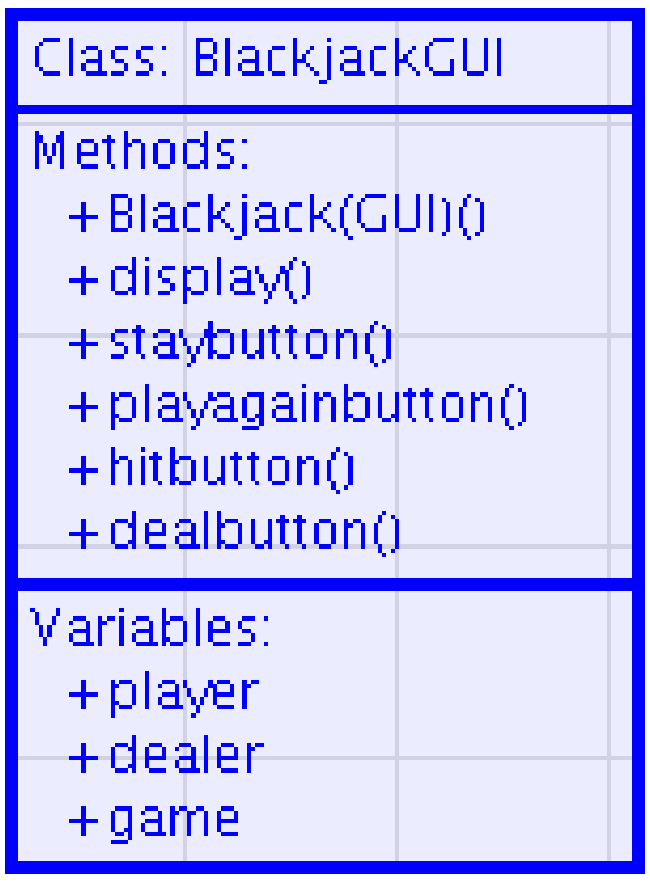

Figure 3: BlackJackGUI class shown in its selected view

When a class object is created, the user is given the opportunity to name the class and provide information about the variables and methods. The class is given a default name if one is not supplied. Double clicking the class object reopens it for editing. The text areas used in the class (class name, methods, and variables) support cut/copy/paste and drag and drop. Figure 4 shows a class being edited. Movie 1[1] shows an example of how a class is created.

[1] The movies can be found at http://perez.cs.vt.edu/minimuml/movies/movie<#>.mov

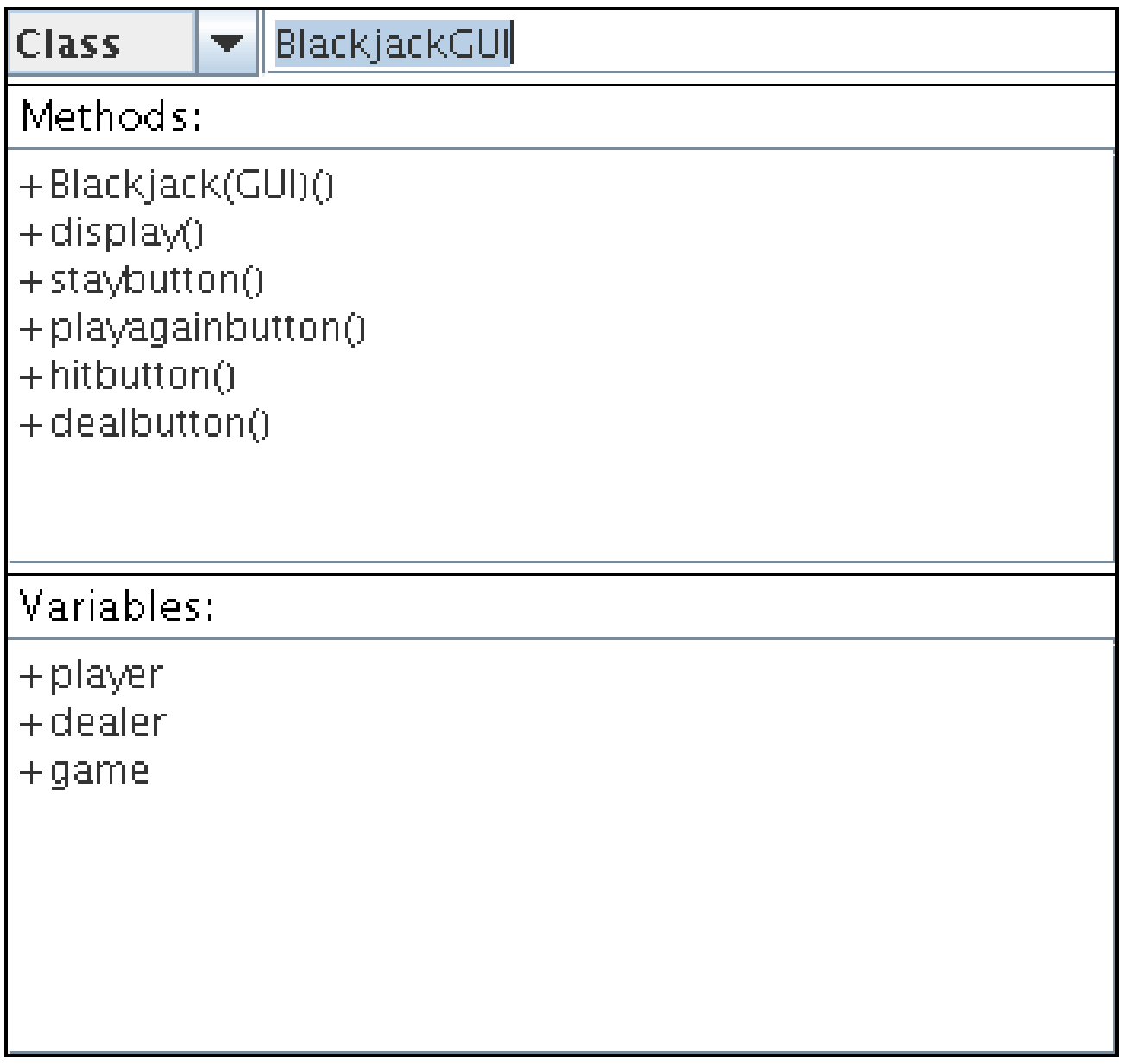

Figure 4: BlackJackGUI class as seen in edit mode

### 4.2 Connections

Connections are simple arrows connecting two classes. These are dependent objects and must have a source and a destination. As such, the deletion of one of those classes deletes the connection as well. In minimUML, connections only have semantic meaning when they join two objects, so, by only allowing that situation, we improve the user's ability to create correct diagrams. A connection can be given four values: generic, association, aggregation, and inheritance. The position of the label and the line itself is automatically determined and is updated as classes are moved about. As connections are drawn, they default to the generic value and can be changed to different values later. This may benefit novice users who may understand that a relationship is needed, but not the type that is needed. Our approach is opposed to a style used in many other UML diagraming tools where separate connection objects are provided for each type. Besides cluttering up the interface with extraneous buttons or menu options, it seems to be very unnecessary. The differences between these objects are their appearance and a little internal state. Separating them and, frequently, not permitting conversions between the various types makes little sense. This may be useful for those who know exactly what they are doing, but does not help those who are trying to explore their ideas. The approach used by minimUML allows for a dependency to be defined and then let its nature be determined and redetermined as the design progresses.

Connections are created by simply selecting the connection tool and dragging from the interior of one class to the interior of another. Figure 5 shows a connection between two classes. Like the class object, double clicking the connection allows the object to be edited. This is accomplished through a simple dropdown list. Figure 6 shows a connection while its type is being edited. Currently, there is no way to change the direction of the connection, the student would simply have to delete the connection and create it in the other direction. Movie 2 shows an interactive example of how a connection is created.

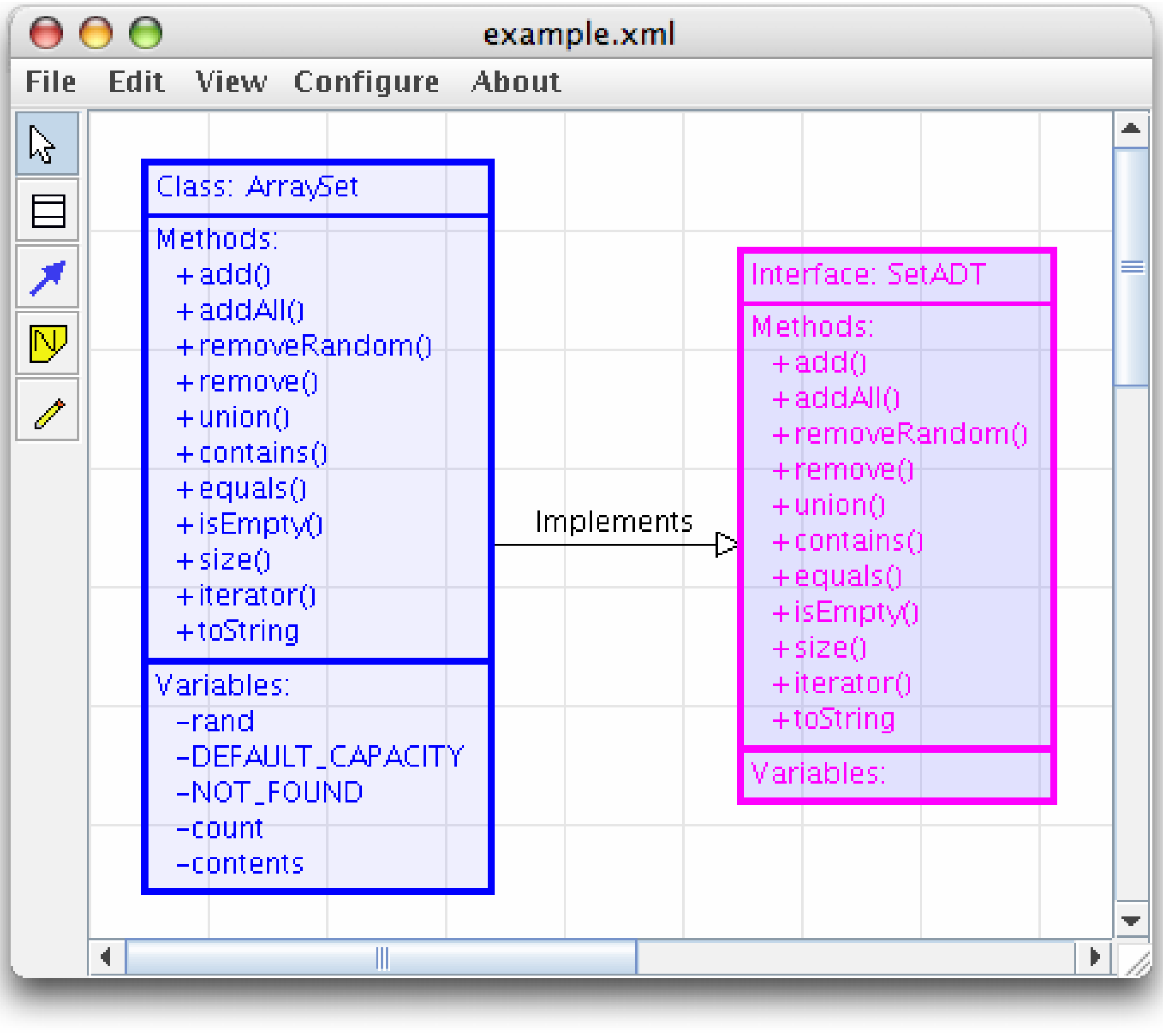

Figure 5: Connection between Classes

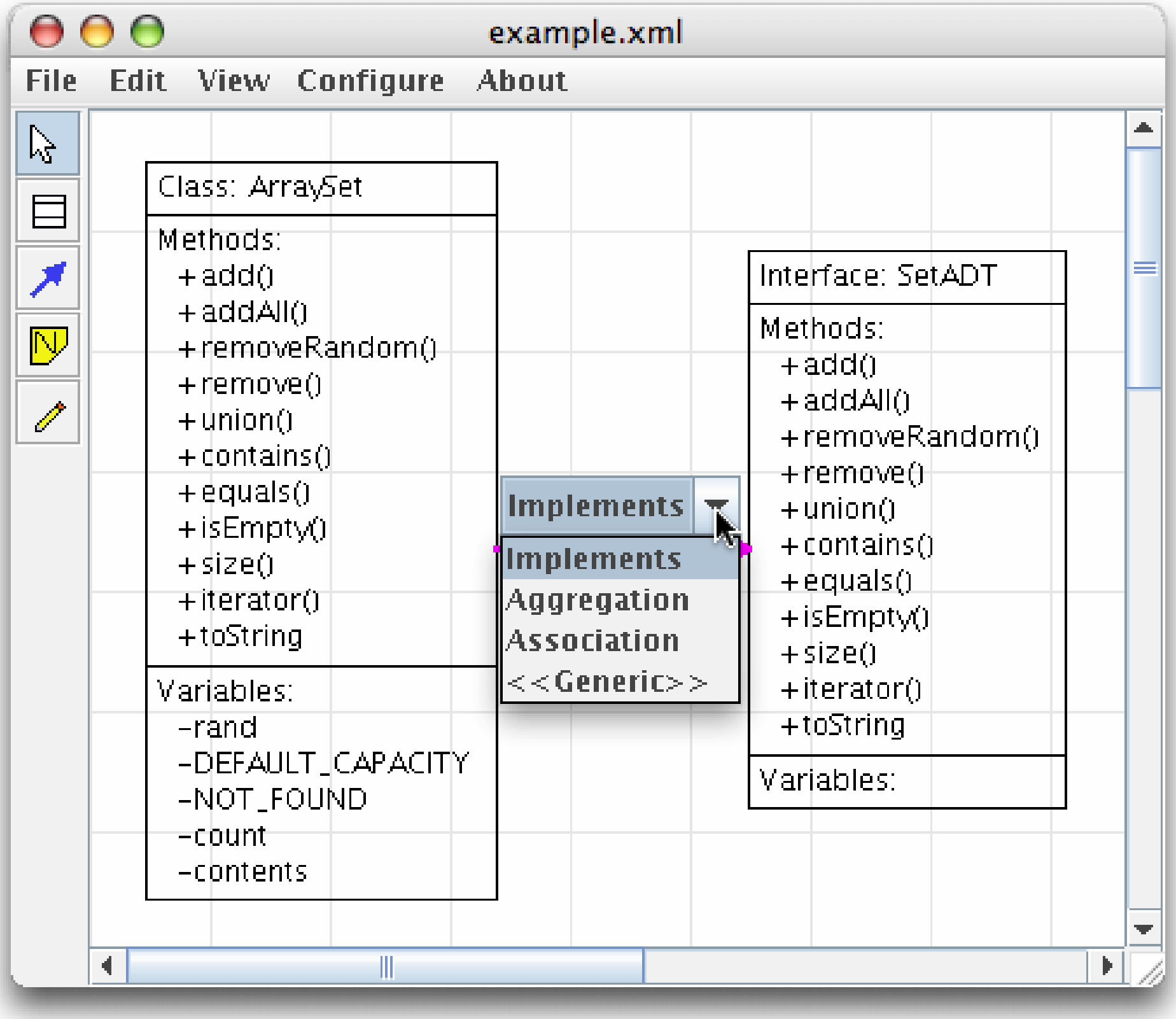

Figure 6: User editing of a connection

### 4.3 Annotations

minimUML supports two types of annotations: electronic sticky notes and freehand drawings, referred to here as glyphs. The electronic sticky notes are areas of typed text that automatically resize to show its contents. These two annotation devices provide a large amount of flexibility to minimUML. Graders can mark up diagrams using a keyboard or a pen device. Students can use the annotation tools to complement tool functionality by typing or drawing in the needed information.

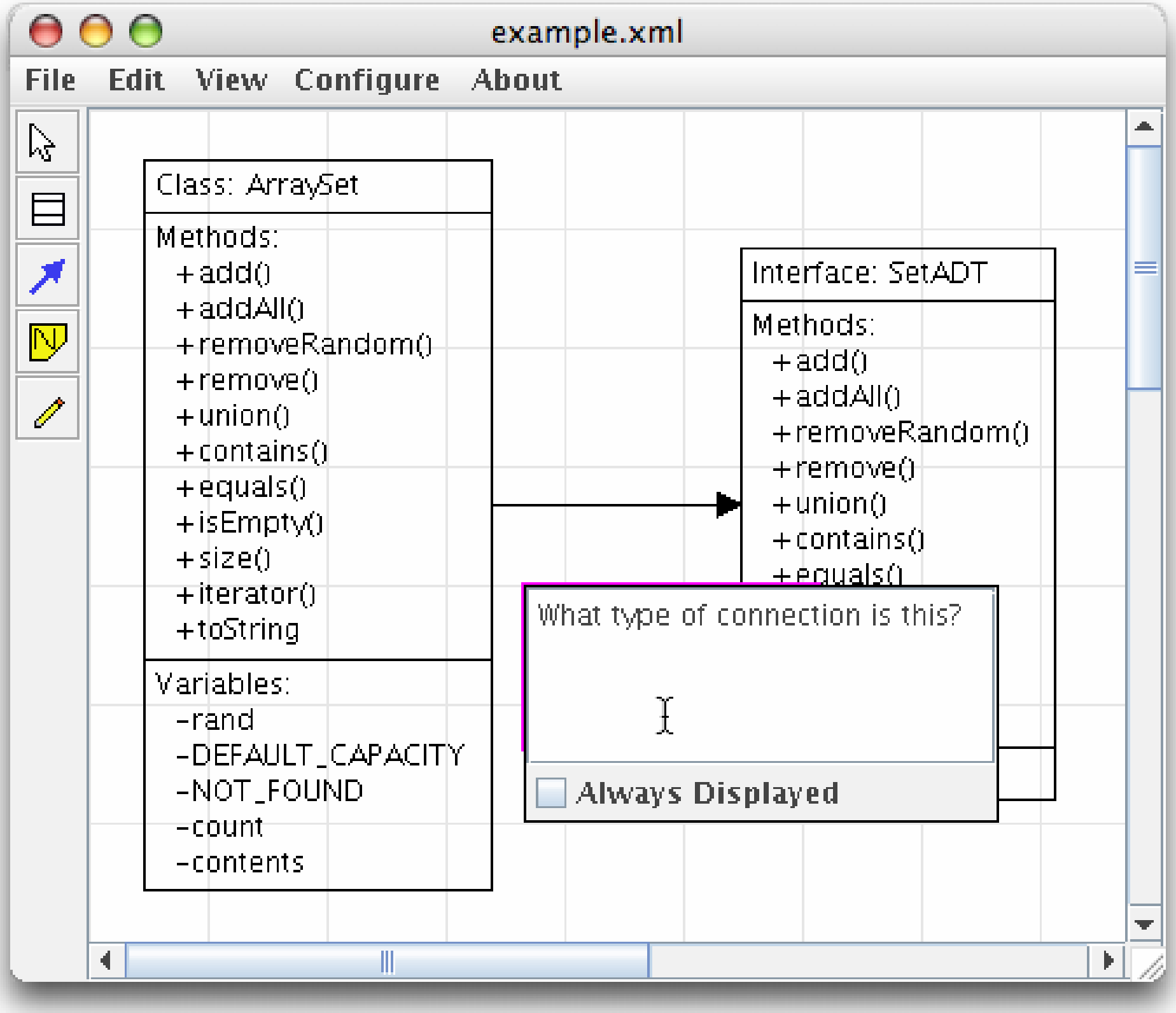

Figure 7: Creating an electronic note

Sticky notes are created in a manner similar to classes. Choosing the note tool and clicking in the diagraming area presents the user with a text area where an annotation can be typed (see Figure 7). By default, a sticky note is displayed as an icon in the diagram until it is selected or moused over, but they can be modified so that they are always shown (see Figure 8). Holding down the ALT key (option key on the Macintosh) expands all of the note icons, to permit easy browsing of the notes. Sticky notes can be edited and moved around to different locations in the UML diagram. Finally, the text area supports cut and paste and drag and drop.

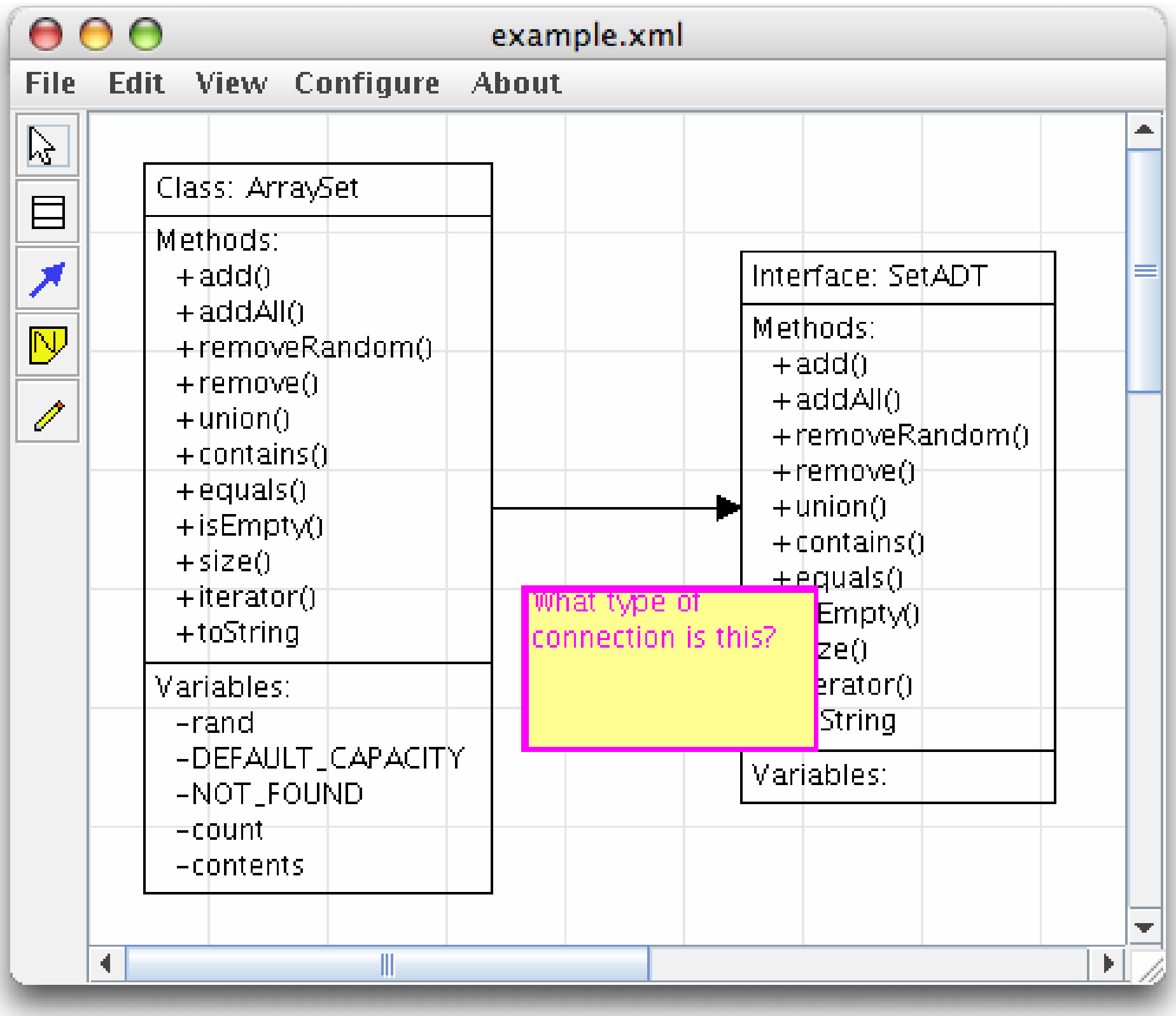

Figure 8: An electronic note shown in its expanded form

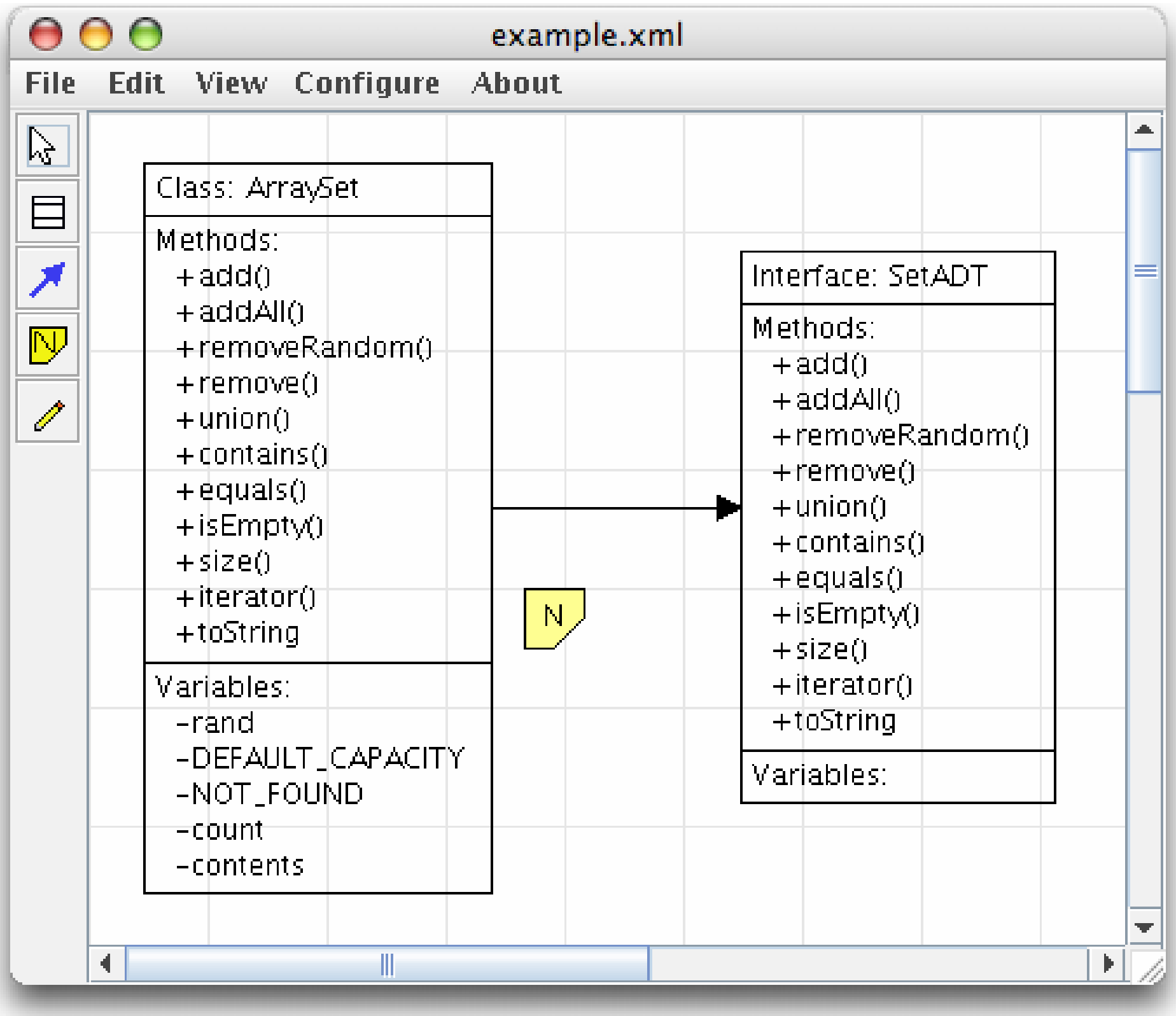

Figure 9: An electronic note shown in its collapsed form

Glyphs are freeform annotations intended to mimic pen-based markings but done with the mouse. They are designed to support any kind of annotation that may be required, such as striking out or circling a section of a UML diagram (see Figure 9: An electronic note shown in its collapsed form
).

Glyphs are made by dragging the cursor across the screen with either a mouse or a stylus. They can be selected, moved, deleted, and otherwise acted upon like any of the other objects, but they cannot be edited. Figure 10 shows comments written with the glyph tool. Movies 3 and 4 show these two annotation tools at work.

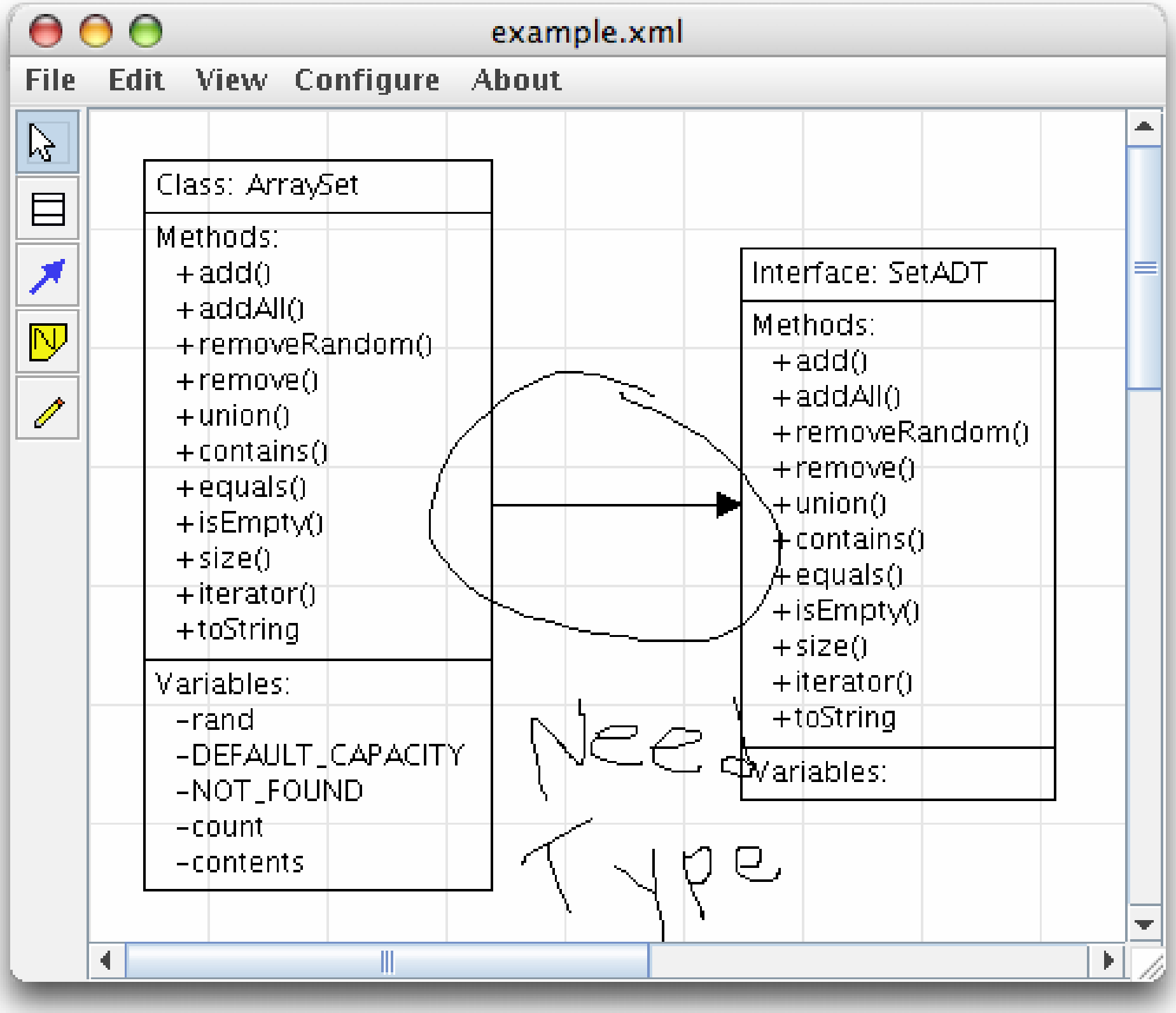

Figure 10: Use of glyph to provide comments

### 4.4 Undo/Redo

In addition to the four diagraming objects, minimUML provides a number of other useful functions. In an effort to support exploratory learning, the tool provides undo/redo capabilities for all actions. There is no set limit on the depth of the undo stack, so a user can go back to the point when a window was first opened.  This feature is associated with a window rather than the current data file, so it allows the undoing of the loading of a file. Standard shortcut keys are associated with the undo and redo actions. Movie 5 shows an example of how the undo command works. When a class is deleted, the connections to and from that class are also deleted. When the deletion is undone, even the connections are restored. An example of this is shown in Movie 5.

## 4.5 Clipboard Support and Drag & Drop

Clipboard support (Cut/Copy/Paste) and drag and drop can be used to move objects among minimUML windows and other windows in the operating system. The standard keyboard shortcuts are supported for the cut/copy/paste commands.

In the basic form of clipboard support, minimUML supports copying of any selected object (except connections) and pasting it in a different location in the same window or in a different window. When multiple classes are copied, all the connections among the classes are copied too. Movie 6 shows an interactive example of several classes being copied to another minimUML document. The movie shows how connections are automatically copied when its two connecting classes are copied.

minimUML makes use of Java's DataFlavor classes, thus providing support for multiple data types in the clipboard. Classes copied or cut into the clipboard can be pasted in other applications in either text form or graphical form. Pasting data from minimUML into a text editor (see Movie 7) produces a Java source code equivalent of the UML objects copied (see the next section for more information on code generation from UML diagrams). Pasting data into a graphical editor produces a graphical representation of the UML objects copied (see Movie 8). These provide support to transfer portions of the UML design to other applications for writing reports, grading, or presentation purposes.

Drag and Drop support provides the same functionality as that provided by the clipboard support. Classes can be dragged from one window to another one within minimUML and the effect is the same as copy/paste. Classes can be dragged to a text editor window with support for drag-and-drop and the source code for the classes will be inserted into the text editor. Classes can also be dragged into a drawing application and inserted in an image format.

## 4.6 Code Generation

In order to support the transfer of UML designs to development environments, minimUML supports simple code generation in C++ and Java. This is meant to provide a stepping off point for actually implementing the design rather than a way to produce reliably compilable code. There is no validation done on the syntax of the variables or methods. We include this as a way to help students visualize how their diagrams would actually be realized in code.

Code generation is available in three ways. First, as an export of the full design. The student can export the whole design as C++ or Java. Second, parts of the diagrams can be pasted into text editors. Third, parts of the diagram are converted to code as a result of a drop action into a text editor that supports drag and drop. The last two support only Java. However, the choice of using Java over C++ was purely arbitrary. In a future version of the software, this will be controlled via a user specified preference option.

## 4.7 Printing

The printing abilities of this tool are surprisingly flexible. Page boundaries can be shown in the diagram and can be oriented in landscape or portrait. The entire document can be printed or just an arbitrary rectangular subsection of the diagram. To produce an overview, the whole design can be printed as a single page. As an aid for reconstructing the printed diagram, a small key at the bottom of each page shows its relative position to the other pages. Figure 11 shows the page boundary shown for a large diagram. If the document were to be printed at this stage, only the portion within the darker rectangle would be printed.

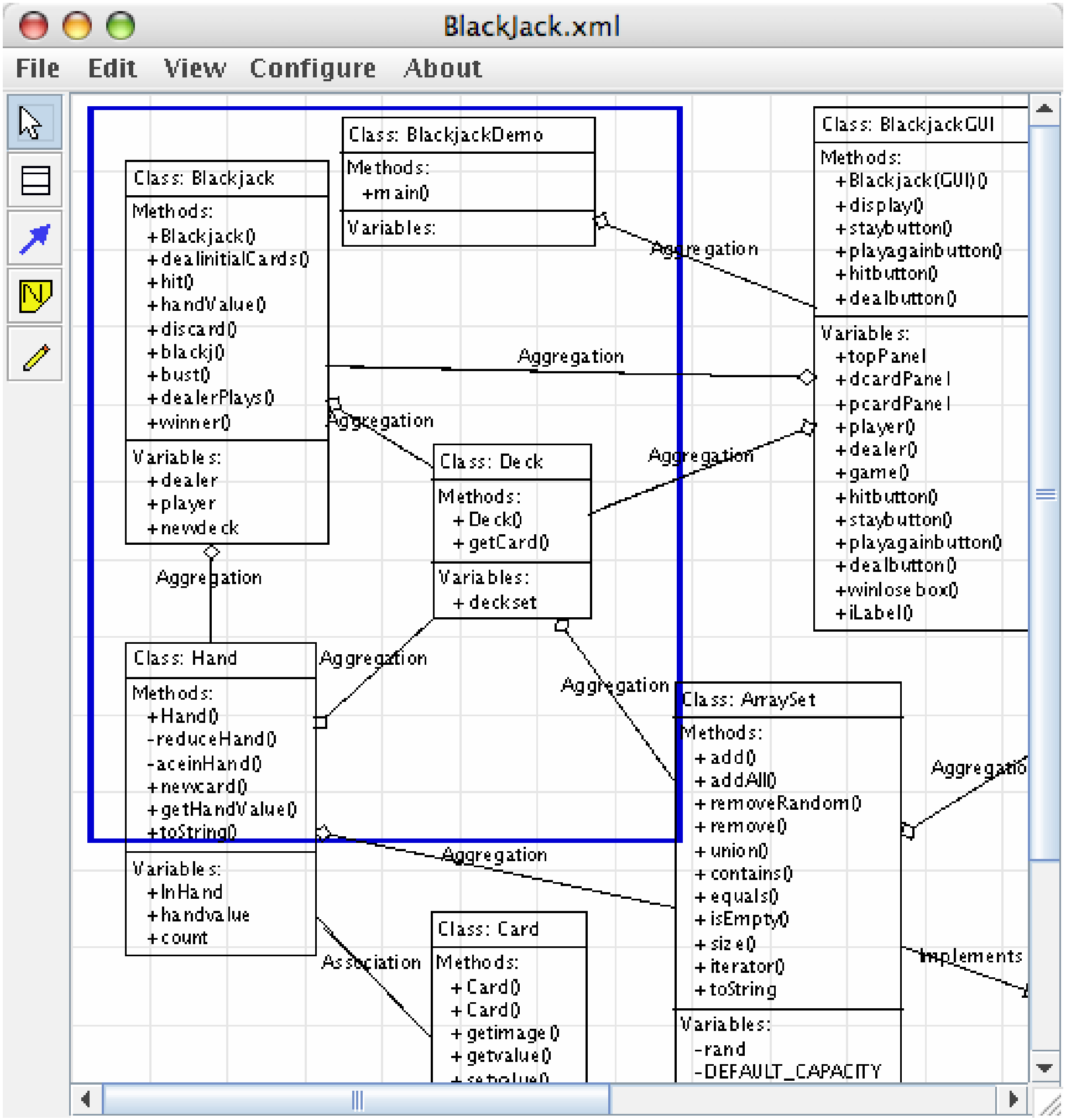

Figure 11: Printing a Selected Area

## 4.8 Other Features

Finally, we've included a number of simple, but sometimes forgotten, features to make this a more useable product. Multiple selection is supported. This can be done by dragging a selection box across the diagraming area. The CTRL key (shift key on the Macintosh) can also be used to toggle the selection of individual objects. Zooming is allowed in increments from 10% to 500% (see Movie 9 for an interactive example). Dragging a minimUML file into a minimUML window will open a new window with the contents of the file.

## 5. EVALUATION OF MINIMUML

We used two different ways to evaluate minimUML. First, we evaluated the general usability of the tool. This was conducted using a typical usability evaluation with a small number of participants. The second phase of the evaluation was done in a classroom where students used the tool to prepare UML designs. We asked the students to fill out a satisfaction survey to assess how easy it was to use UML without prior training. This section describes both evaluations and the results obtained from them.

## 5.1 Usability Evaluation

To validate the design of our tool, we ran a simple usability study with six participants. The participants were Computer Science majors at Virginia Tech. They ranged in age from 20 to 22 and they were an even mix of juniors and seniors. All were male. Most of the students had little UML experience. Three rated themselves as beginners, two as intermediate users, and one said he had no experience. They were given a small amount of class credit for their participation.

To begin the session, the students were given a very brief introduction to the program. This introduction was intended to be similar to one that a typical instructor would give in a lab session the first time students were given the tool. The participants were then given a couple of minutes to become familiar with the tool's interface. When they finished exploring, they were given a simple task description detailing a movie rental system that they needed to diagram [See Appendix A]. The problem was specifically designed to provide situations for the use of aggregation, association, and inheritance. A reasonable solution to the task requires approximately ten classes.

The problem used is very typical of first-year design problems; it resembles, in size and complexity, a number of case studies from introductory textbooks. Horstmann (2004) provides an example problem about a voice mail system (page 63). All told, the UML diagram presented contained six classes. Lewis and Chase (2005) give three additional examples of these types of design problems. One concerns a Blackjack program with a nine class diagram. Another diagram for a digital calculator had eleven. Finally, there was a webcrawler consisting of 18 classes. The general level of difficulty is also roughly equivalent to our design problem.

In the interest of time, the users were asked not to worry about adding any accessors or mutators to the diagrams. They were given 30 minutes to complete the exercise. At the end of the session, they were given a questionnaire [See Appendix B].

Since one of the aspects being measured here was the ease at which the interface was learned, we deliberately gave them cursory instructions and limited the time they spent looking at the tool before starting the task. We attempted to create a situation similar to one in a typical lab session for an introductory course. In addition, they were told that they could not ask questions about the program during the session.

## 5.2 Results

Five of the six students took the full 30 minutes to complete the task. The sixth student finished in approximately twenty minutes and spent the rest of the time exploring some of the tool's other features. The response to the program was generally positive. While there were a number of issues raised, the participants found it very easy to use and seemed to be comfortable using the program.

As the students created their designs, there were a number of mistakes observed as they became familiar with the interface. As discussed later, there was some confusion over the meaning of buttons and how to interact with the objects. Despite this, everyone finished the task and no one appeared to have major issues learning the interface. This is encouraging as it suggests that it is reasonable to introduce the tool to students and to have them use it in the same class or lab session.

On the questionnaire, questions 3 to 23 were 5 point likert scales with 1 being strongly disagree, 5 being strongly agree, and 3 being neutral. Questions 15, 20, 21, and 23 were worded negatively but have been inverted in the data analysis for consistency. [See Appendix C for full results.]

### *5.2.1 Usability Evaluation*

Questions 3 to 10 dealt with the usability of the tool for the task with which they were presented. The participants found it easy to create and modify both the classes and connections with almost all of the ratings being agree or strongly agree for these questions.

They were more ambivalent to the note and glyph tools but that is not surprising as the task did not require their use. When asked about the other features of the tool, specifically undo/redo and zoom, they also responded favorably. Overall, they seemed to agree that the tool was adequate for the task given. Figure 12 shows a graph with the results.

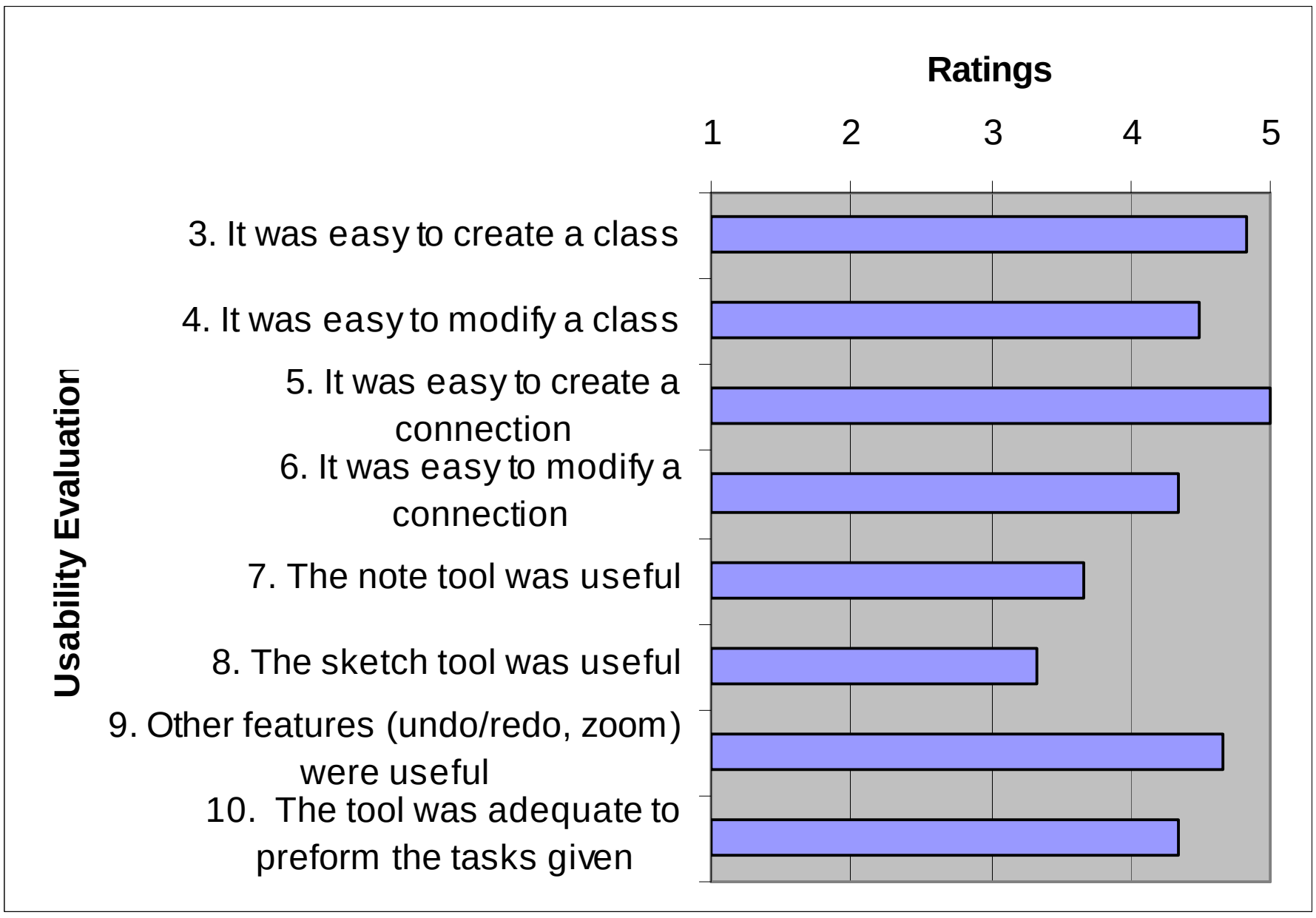

Figure 12: Average ratings for the usability questions

### *5.2.2 User Interface Evaluation*

The responses to 11, 12, and 13 relate to the tool's interface (see Figure 13). Participants found it easy to learn and use, but were less enthusiastic about the amount of feedback provided by the tool. From the other responses in the questionnaire, this may be attributed, at least in part, to the lack of tool tips for the buttons. Several students were observed to be initially confused over the purpose of the buttons in the toolbar. Tool tips may alleviate this problem and have been added to the interface.

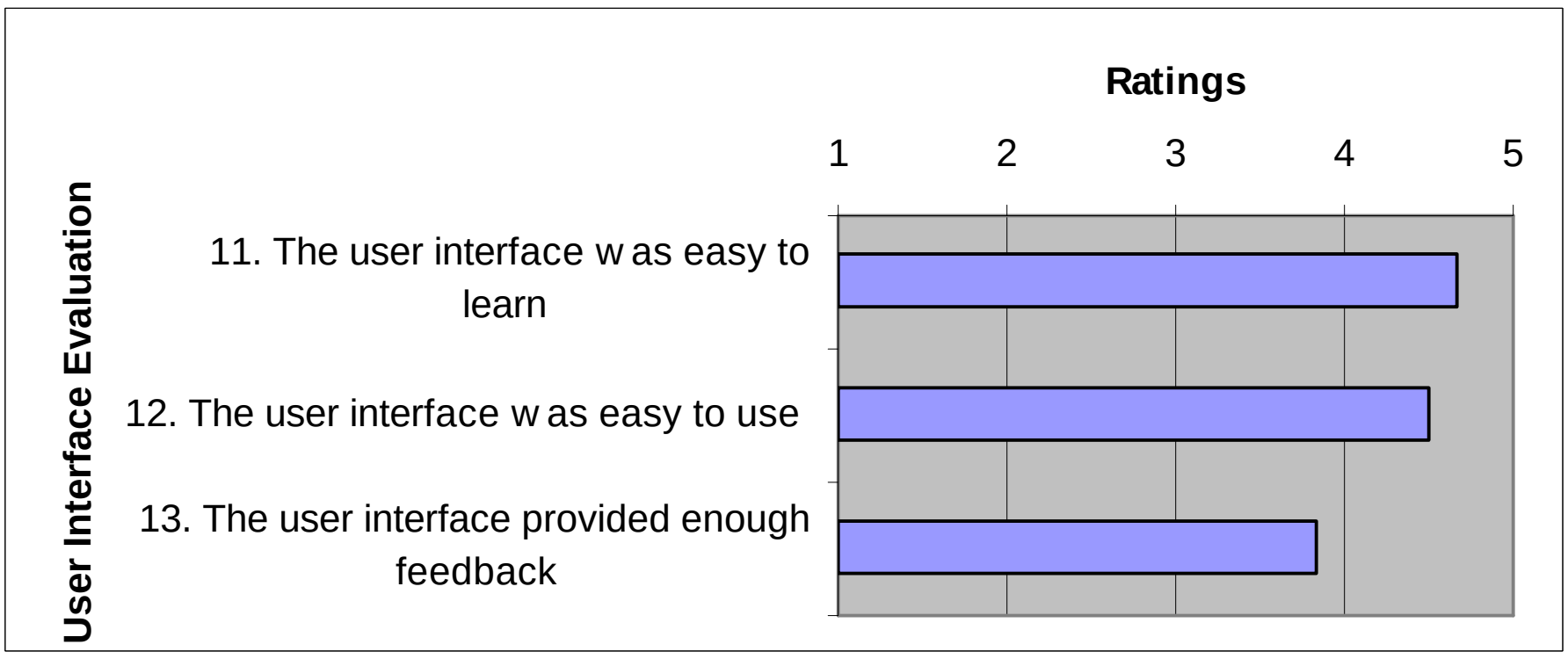

Figure 13: Average ratings for the user interface questions

*5.2.3 UML Support*

The next four questions revolve around the amount of UML provided by the tool and its usefulness. (Note that the scale for question 15 was inverted for the analysis to match the others.) The participants seem to be satisfied with the amount of UML provided and the efficiency of the tool but they were fairly neutral about whether more UML functionality was needed. They also only moderately agreed that the tool was useful to them. In part, this data is skewed by the inclusion of a student who, judging from the UML tools he listed as having used and from the nature of his comments, appears to have significantly more experience with UML than the others. Since this tool is intended for use in introductory level courses and was designed as such, it is no surprise that people with more UML knowledge would not find it useful. It is more important that the target audience, those with less background in UML, be satisfied with it. It should be noted that the task did only require the amount of UML provided by the tool. So, it is not unexpected for the responses to fall as they did. It would be more worrisome if the ratings were more negative as they would show a serious flaw in the tool. Figure 14 below shows the results of this section.

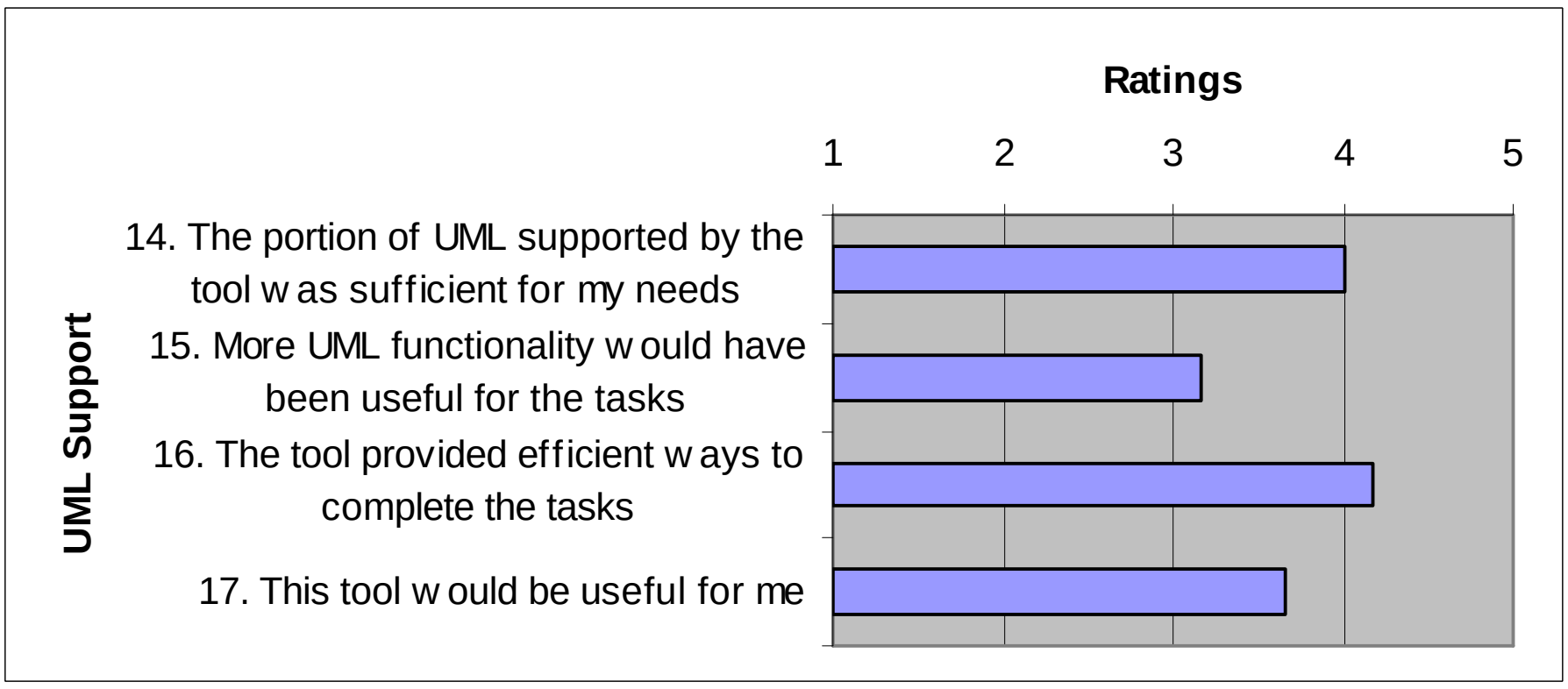

Figure 14: Average ratings for the UML support questions

*5.2.4 Exploration Support*

The final questions, 18 to 23, focus on the exploration of the interface (see Figure 15). (Note that the scale for questions 20, 21, and 23 were inverted in the analysis to match the others.) The students reported being comfortable exploring the interface without negative consequences. However, they were more uncertain about the outcome of their actions. This may be related to the perceived lack of feedback from the tool. They were also only somewhat in agreement that the interface did not limit their actions. Again, this is partially biased by the one experienced participant who felt very limited by the software. The fact that the rest did not feel restricted indicates that this may not be much of a problem. What is of more concern is that they were close to neutral on the difficultly of making errors. This concern is, in some respects, mollified by the positive responses to the ease of error recovery and to the perceived lack of lasting consequences for mistakes. Although there is not as much error avoidance as may be desired, the facilities for resolving those missteps seem to be adequate. As a whole, it seems that the tool permits exploration fairly well.

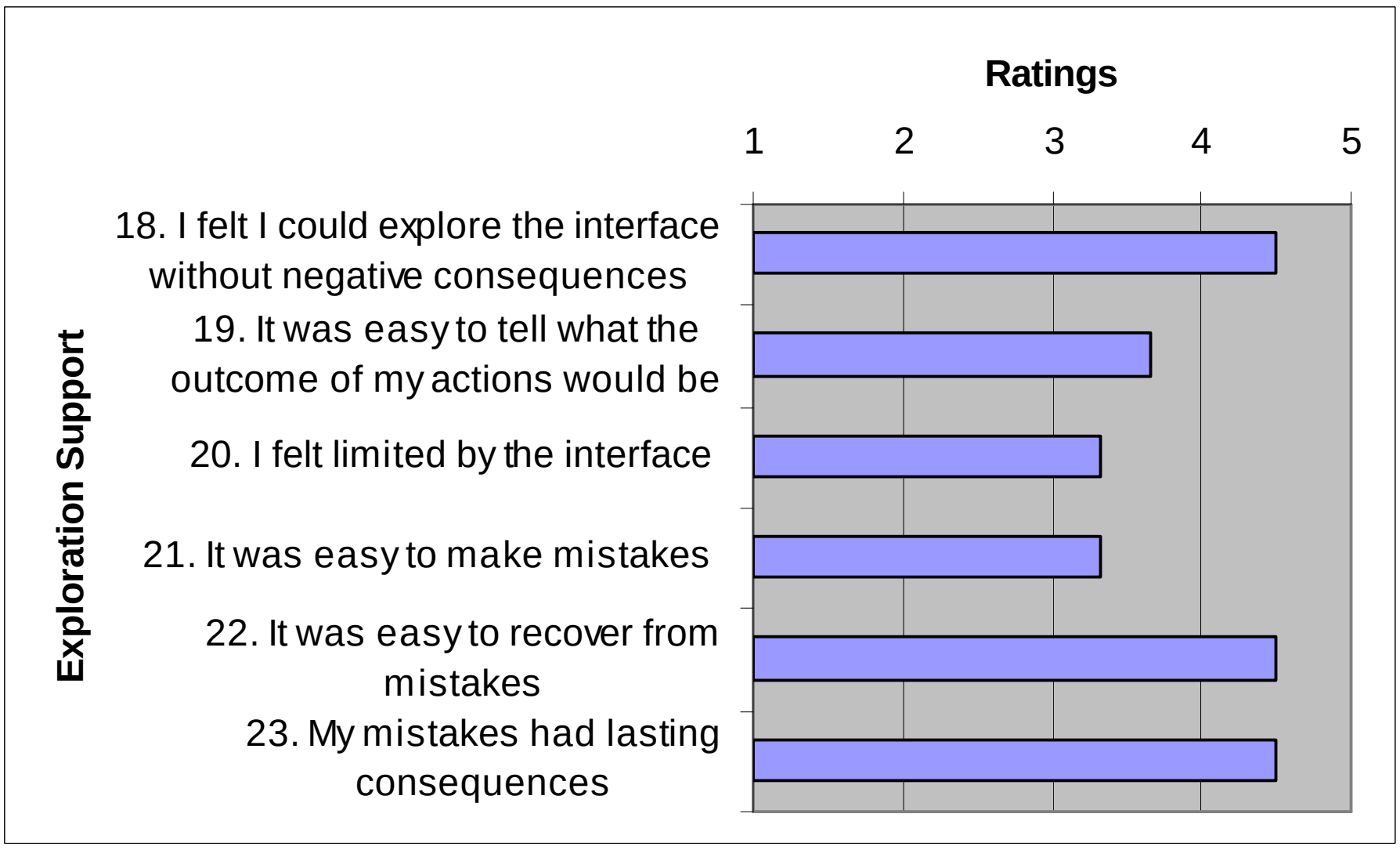

Figure 15: Average ratings for the exploration support questions

*5.2.5 General comments*

After the likert scale questions, the participants were asked to describe what they like and disliked about the classes, connections, and the other features of the tool. In addition, they were asked for their overall impressions and what features were missing and should be added. For the classes, they were almost unanimous in saying that they liked the simplicity of creating and modifying them. One person liked the fact that the class name, variables, and methods were all immediately available to him. On the negative side, one user wanted to be able to manually adjust the size of the box and we observed several attempts to resize classes. We chose to use automatic sizing to ensure that each class displayed all its information and that the information was legible. With manual sizing, there is a problem where shrinking a class can obscure information or, if the font size decreases with the class size, render the text unreadable. Since there does not seem to be much advantage in increasing the size of a class, we disallowed it completely.

Another student wanted to be able to specify the type of class, such as abstract or interface. In response to this comment and to the perceived need for interfaces for certain kinds of designs, support for interfaces was added. When editing a class, the student is presented with the option to select whether the object is an interface or a class. While there is no explicit support for other qualifiers, such as virtual, static, or abstract, a simple solution would be to add any such information to the class name.

For the connection objects, the students again said that they were very easy to create and use. One person did not like that connections were created and then given a type. He would have preferred separate tools for each type of connection. We have discussed our reasons for using our method earlier. It is interesting to note that the student wanting this distinction was the one with the most UML experience. This supports our claim that more knowledgeable designers may think of the relationships between classes in concrete terms while beginners may recognize it as a relationship but not necessarily know what kind. He also expressed dislike for our nonstandard method of labeling the arrows; we used text labels versus the standard diamonds, triangles, etc. Since this is intended for novices, we used the text labels in order to avoid any confusion over the meaning of symbols in the diagram. A better approach would have been to use a combination of the symbols and the text labels. That removes the ambiguity of the symbols for novices while adhering to the standards. minimUML has since been updated to implement this.

A more serious problem we discovered with the connections was that five out of the six students did not realize that they could be edited. In fact, several of them suggested that we add the ability to specify the type of relationship; they did not realize that double clicking the connection would allow just that. It seems that the problem stems from having the connection default to a generic relationship. In this state, there is no text label shown. Thus, it appeared to most of the users as a simple line that did not have an associated, changeable value. To solve this, we decided to have the generic connection display a value when it is highlighted or selected. This change in state, as the line is manipulated, should provide enough of an affordance to indicate that it may be edited.

When asked about the features that they liked, most replied that they liked the ease at things were created or manipulated. One noted the cut and paste abilities and the flexibility of the glyph tool. Several people mentioned the perceived lack of a way to add a type to the connections as an aspect that they did not like. Overall, the participants said they found the tool useful and easy to use.

Finally, the participants suggested a number of ways to improve the tool. One addition mentioned by a couple of people was the addition of tool tips. Another suggested a popup edit menu that allows the user to switch mode as an alternative to using the toolbar. Both are simple and effective ideas for improving the usability of the program. Another suggestion was to create another view of the diagram as a list of the classes. While this would certainly be useful for larger projects, we are not sure how important it would be for the size of diagrams for which this tool is designed. Its value may not justify the screen space it takes up. One solution would be to allow the user to show or hide the list at will or to place it in a separate window. At this time, this has not been implemented.

### 5.3. Classroom Evaluation

In addition to our usability study, we used minimUML in the lab sections of a first year object-oriented course for the first time in the spring of 2005. This section reports on our experiences of that first use.

Students were asked to use the tool in a 2 hour lab session. They were given a brief introduction to the tool and asked to prepare a small design. As the end of the lab, we asked the students to take a survey about the ways they created their diagrams and the amount of time they spent learning to use the tool. We received 34 responses to the survey.

Of the 34 students that completed the survey, 29 were freshmen, 4 were sophomores, and 1 was a graduate student. All but one were 18 or 19 years old; the other participant was 23 years old. The majors of the students were also very homogeneous. There was a single math major among the computer science students. Only one female responded.

Only a very few of the students reported that they had much experience with UML. Half said that they had no experience with UML at all. Another 41% classified themselves as beginners who had used UML in some small class projects. Of the remaining three students, two were of an intermediate level and one was an expert.

First, the students were asked about what approach they used while designing classes. A majority of the students (71%) said that they frequently or constantly modify their classes as the design progresses. Some of these changes are due to iterative approach that some students used when adding methods to their classes. While almost half of the students (47%) stated that they completely define their methods from the start, 38% only entered some of the information for each method, such as a return type and/or some of the parameter information. One person reported initially giving methods only a name and three others used different strategies depending on the situation. This highlights the need to allow students flexibility in the manner in which they design. Enforcing syntactically correct methods at this stage would have negatively influenced half of the class.

The students dealt with the variables in a more uniform manner. Most (79%) always supplied a name and a type for each variable. Three students mentioned that they just put in a name while one person gave only the type. Two people said that their approach depended on the situation. While there is not as much variance in the approaches as there was with the methods, there is still enough to argue for few restrictions on how values are entered.

Next, the students were asked about how they made connections between classes. Over half (59%) of the students reported that thy made connections between classes and selected a type later. In addition, 35% said that they change the type of the connection during the design process, sometimes frequently. This behavior supports our rationale for providing a connection object that can switch between types instead of providing separate connection objects. It directly provides for this action while the other method requires connections to be deleted and then recreated.

Finally, we asked about the time that the instructor or teaching assistance spent explaining how to use minimUML and how much time the student spent exploring it. 32% of the students responded that they received no instruction on how to use the program. In additional 15% said that they had less than a minute introduction to the tool, 24% said 1 to 5 minutes, and 21% said 5 to 10 minutes. When asked if that was enough time, 56% said that was sufficient time spent and, somewhat surprisingly, 29% said there was too much time used. A third (33%) of the class remarked that they were comfortable using the program with that much instruction, even if they had some problems with it. An additional third (38%) reported that they were neither uncomfortable nor comfortable with the tool. Of the remaining students that did not feel comfortable, only one person said that they had many problems using it. For the instructor, it would be reasonable to spend between 5 and 10 minutes demonstrating a new tool to a class at the beginning of a lab session or a lecture and it would seem that minimUML is simple enough for this to be adequate time for the majority of the students.

The time that the students spent exploring the program on their own followed a similar pattern. A quarter (24%) of the students immediately started using it for their lab assignment. 15% played with the program for less than a minute, 32% used 1 to 5 minutes, and 21% spent 5 to 10 minutes. Only a single student reported spending more than 10 minutes learning how to use the program. Again, most students thought this was sufficient (47%) or too much time (35%).

Our survey of students using minimUML as part of a lab has shown us that it supports some of the basic behaviors that the students engage in. It allows for the iterative and sometimes ill-defined nature of their designs and it allows them to easily manipulate the connections as their concept of their design changes. In addition, it would seem that minimUML can be learned and used effectively within a lab session in fairly little time.

## 6. FUTURE WORK

While minimUML supports several methods for exporting designs, either to paper or to another electronic format, it does not currently allow for electronic submission to other tools, such a course management system. Ideally, there would be a way, similar to the method used by Allowatt and Edwards for their Eclipse plug-in (2005), for students to submit and retrieve files directly from the tool. These diagrams, once submitted, could then be sent to teaching assistants who would assess them and upload graded versions. Alternatively, they could be redistributed to the class for a peer review exercise. In either case, the ability to upload and download diagrams directly through the program would streamline the process and allow the students to focus on their designs rather than on file management. We are currently exploring how to implement this feature.

We are also looking into a simple and generic method of downloading diagrams into minimUML. One possibility is to allow the user to drop URLs into the work area to open those files. While this would not allow the students to upload files, it does not require server support to handle downloading of files.

This program would also benefit from the ability to group and ungroup objects. Some objects, notably a group of glyphs used to write out a sentence, are logically a group and it would be convenient to act upon them as if they were a single object. This would allow for the creation of locational relationships between objects and would make the diagram easier to manipulate in some cases.

Being able to import code as a diagram would be another useful feature for minimUML. Students would be able to compare original designs to what was actually coded in their programs. With a little introspection, they can learn a great deal from the differences in the diagrams. For instructors, this may provide a way to gauge how well the students are learning to design or it they are slapping together a diagram and jumping right into the code. Either way, it would supply very useful information. Providing some form of automatic diagram comparison to aid with these kinds of tasks would be a possibility as well.

If students were able to import their code as a diagram, an interesting idea to pursue is a form of round trip editing for their code. Alphonce and Martin have implemented this kind of functionality in their Eclipse plug-in Green (2005). Ideally, a student would make a design and then export it to code. After that point, the diagram would reflect the changes in the code and vice versa. This would give the students two views of their work with minimal cost to move between them. That being said, there are difficulties to overcome to make this idea work. Synchronizing the data between the views, so that work is not lost, is a considerable problem. This is especially an issue if the student wants to work with both views at once. Handling some of the intricacies of the code with the limited amount of UML provided by the tool is another concern that would need to be explored.

Finally, adding in some "style commenting" features may be useful. At Virginia Tech, researchers have been successfully using an automated grader called Web-Cat (Edwards 2003), which uses, among many other tools, an open source style checker that analyzes naming conventions and performs other static checks on code. By integrating a tool such as this one into minimUML, we could comment on the students' style before they start writing code. For example, we could check variable and method names to ensure that they are not too short. Or, we could check the connections between classes to ensure that situations, such as circular inheritance, do not appear. This might reinforce good programming style, even when they are working at an abstract level. We are exploring how to incorporate such tools into minimUML, and if their benefit would outweigh the integration effort or the added complexity in the interface.

While minimUML works well for the most part, it does have some known flaws. One issue arises when there are a large number of objects in the diagram. Our original design did not include glyphs. Thus it was assumed that in its intended use, diagrams would have perhaps a couple dozen objects and very large designs may have 50 or 60. This assumption has held true so far in the use that the tool has received. However, the addition of glyphs radically changes this assumption. As we discovered later, a grader using glyphs for annotations created around 250 glyphs for a diagram originally consisting of 11 objects. In this case, the algorithms currently used to detect focus and selection run rather slowly. These should be replaced with more appropriate algorithms and data structures to account for the increased number of objects.

## 7. CONCLUSIONS

In order to support object-oriented design early in the curriculum, we have identified a number of requirements that a good tool should support. It must be easy to learn and use, have features, such as undo/redo, that support exploratory learning and error avoidance, provide flexible printing options, support some code generation, allow for annotations, and focus on abstract design rather than coding. We have surveyed several available UML diagraming tools and found that they did not meet these requirements to our satisfaction. In response, we created minimUML that tries to meet these requirements. We evaluated minimUML and results were positive and support our claim that the tool is appropriate for early CS education.

The goal for minimUML is to provide just enough UML to support learning in the early courses of Computer Science education. The subset used is the same one used in many introductory Computer Science courses. With just class and connection objects, it gives adequate power to create reasonably complex designs while not cluttering the interface nor burdening the user with extraneous features. This helps makes the tool easy to learn and use. Since there are very few restrictions on the way in which the classes are designed, it allows the users to design at an abstract level. There is nothing to prevent them from starting with just a few high level details which they can expand upon sometime later. Furthermore, since it is so uncomplicated, it is hard to make serious errors. Common, and expected, features, such as multiple selection and cut and paste, enhance its usability, while the undo/redo allows for exploration. We have taken a minimalist approach to create a UML design tool to make a simple, but useful, tool.

## ACKNOWLEDGMENTS

We would like to acknowledge Jaime García-Ramírez who has taken up the job of updating the minimUML tool. We also would like to thank the students in CS 1705 and CS 1706 at Virginia Tech for graciously letting us explore the use of new tools that will benefit future students. The last stages of this tool were developed with support from Microsoft Research University Relations under their Tablet PC and Computing Curriculum Program.

APPENDIX A

Movie Rental Database

MegaMovies Corp. Inc. Ltd. has hired you to create a new program for their movie rental database. After giving you the details of what they want for their software, they ask you to produce an initial design that will show them how the program will be built. (You've noticed that their concept of the database is missing a lot important points, but you'll bring that up when you show them your design. And then ask for more money.)

The database consists of two tables, Movies and Customers, and their associated indexes. For the Movies table, MegaMovies wants it searchable on the movie title, the unique MegaMovies' id number, and any of the actors/actresses that star in the movie. The Customers table should allow lookups on the customer's name and the customer's id number.

For each movie, there need to be some standard information, such as the release date and the producing studio, and it is also necessary to store information about all the store's copies of a particular title and all of the reviews about the film. Each movie copy must have a unique id, some information about its current status, and, if it is rented, who rented it.

For each customer, an address and an account balance are needed. In addition, it must be possible to determine which videos a customer currently has rented. Once the basic functionality has been implemented, MegaMovies wants to allow users to write reviews about the movies they've rented, so you need to design a way to connect the customers to the reviews they have written.

Each movie can be reviewed by a critic, by the studio, or by customers. Each review has a reviewer, a rating, and a text review. The studios also provide some text describing promotional events for the movie and that needs to be stored as well. Unfortunately, this information is transmitted in different ways. The critics' critique comes in the form of a small XML file for each movie. The studios mail a CD with all the data stored in a large table and the customers' views are inputted as plain text. Luckily, you already have a XMLFileClass and a StudioReviewTableRowClass that will help with some of the reviews, so you don't need to worry about implementing or designing them.

With all that in mind, set out and diagram an initial plan for the program in UML.

APPENDIX B

## minimUML User Evaluation Questionnaire

Participant Number: __________
Demographics
Gender: Male Female
Age: ______
Class level: Fr Sph Jr Sr
Major: ____________________

### UML Background

1. How would you rate your level of proficiency with UML?
   None Beginner (small class projects)
   Intermediate (several large designs) Expert (used UML extensively)
2. What other UML tools, if any, have you used?

| **Task** | Strongly Disagree | Disagree | Neutral | Agree | Strongly Agree |
|---|---|---|---|---|---|
| 3. It was easy to create a class | 1 | 2 | 3 | 4 | 5 |
| 4. It was easy to modify a class | 1 | 2 | 3 | 4 | 5 |
| 5. It was easy to create a connection | 1 | 2 | 3 | 4 | 5 |
| 6. It was easy to modify a connection | 1 | 2 | 3 | 4 | 5 |
| 7. The note tool was useful | 1 | 2 | 3 | 4 | 5 |
| 8. The sketch tool was useful | 1 | 2 | 3 | 4 | 5 |
| 9. Other features (undo/redo, zoom) were useful | 1 | 2 | 3 | 4 | 5 |
| 10. The tool was adequate to perform the tasks given | 1 | 2 | 3 | 4 | 5 |

| **Interface** | Strongly Disagree | Disagree | Neutral | Agree | Strongly Agree |
|---|---|---|---|---|---|
| 11. The user interface was easy to learn. | 1 | 2 | 3 | 4 | 5 |
| 12. The user interface was easy to use | 1 | 2 | 3 | 4 | 5 |
| 13. The user interface provided enough feedback | 1 | 2 | 3 | 4 | 5 |

| **UML** | Strongly Disagree | Disagree | Neutral | Agree | Strongly Agree |
|---|---|---|---|---|---|
| 14. The portion of UML supported by the tool was sufficient for my needs | 1 | 2 | 3 | 4 | 5 |
| 15. More UML functionality would have been useful for the tasks | 1 | 2 | 3 | 4 | 5 |
| 16. The tool provided efficient ways to complete the tasks | 1 | 2 | 3 | 4 | 5 |
| 17. This tool would be useful for me | 1 | 2 | 3 | 4 | 5 |

| **Exploration** | Strongly Disagree | Disagree | Neutral | Agree | Strongly Agree |
|---|---|---|---|---|---|
| 18. I felt I could explore the interface without negative consequences | 1 | 2 | 3 | 4 | 5 |
| 19. It was easy to tell what the outcome of my actions would be | 1 | 2 | 3 | 4 | 5 |
| 20. I felt limited by the interface | 1 | 2 | 3 | 4 | 5 |
| 21. It was easy to make mistakes | 1 | 2 | 3 | 4 | 5 |
| 22. It was easy to recover from mistakes | 1 | 2 | 3 | 4 | 5 |
| 23. My mistakes had lasting consequences | 1 | 2 | 3 | 4 | 5 |

## Impressions

24. When working with a class, what did you like the most? The least?

25. When working with a connection, what did you like the most? The least?

26. What feature(s), if any, did you like the most?  Why?

27. What feature(s), if any, did you like the least?  Why?

28. Briefly describe your overall impressions of the tool.

29. What should be added to the tool to make it more useful?

## APPENDIX C

| Questions/Students | S1 | S2 | S3 | S4 | S5 | S6 | Range | Avg | Std Dev |
|---|---|---|---|---|---|---|---|---|---|
| 3. It was easy to create a class | 5 | 5 | 5 | 5 | 5 | 4 | 1 | 4.83 | 0.372678 |
| 4. It was easy to modify a class | 5 | 3 | 5 | 5 | 4 | 5 | 2 | 4.50 | 0.763763 |
| 5. It was easy to create a connection | 5 | 5 | 5 | 5 | 5 | 5 | 0 | 5.00 | 0 |
| 6. It was easy to modify a connection | 4 | 4 | 5 | 5 | 5 | 3 | 2 | 4.33 | 0.745356 |
| 7. The note tool was useful | 3 | 3 | 4 | 5 | 3 | 4 | 2 | 3.66 | 0.745356 |
| 8. The sketch tool was useful | 3 | 3 | 5 | 3 | 3 | 3 | 2 | 3.33 | 0.745356 |
| 9. Other features (undo/redo, zoom) were useful | 5 | 4 | 5 | 4 | 5 | 5 | 1 | 4.66 | 0.471405 |
| 10. The tool was adequate to perform the tasks given | 5 | 4 | 3 | 5 | 5 | 4 | 2 | 4.33 | 0.745356 |
| 11. The user interface was easy to learn | 5 | 4 | 4 | 5 | 5 | 5 | 1 | 4.66 | 0.471405 |
| 12. The user interface was easy to use | 5 | 4 | 4 | 5 | 4 | 5 | 1 | 4.50 | 0.5 |
| 13. The user interface provided enough feedback | 4 | 3 | 3 | 4 | 5 | 4 | 2 | 3.83 | 0.687184 |
| 14. The portion of UML supported by the tool was sufficient for my needs | 5 | 4 | 2 | 4 | 5 | 4 | 3 | 4.00 | 1 |
| 15. More UML functionality would have been useful for the tasks (Scale Inverted) | 2 | 3 | 4 | 3 | 2 | 3 | 2 | 3.16 | 0.687184 |
| 16. The tool provided efficient ways to complete the tasks | 5 | 4 | 3 | 4 | 5 | 4 | 2 | 4.16 | 0.687184 |
| 17. This tool would be useful for me | 4 | 3 | 1 | 4 | 5 | 5 | 4 | 3.66 | 1.374369 |
| 18. I felt I could explore the interface without negative consequences | 5 | 4 | 5 | 4 | 4 | 5 | 1 | 4.50 | 0.5 |
| 19. It was easy to tell what the outcome of my actions would be | 4 | 4 | 4 | 3 | 2 | 5 | 3 | 3.66 | 0.942809 |
| 20. I felt limited by the interface (Scale Inverted) | 2 | 3 | 5 | 3 | 1 | 2 | 4 | 3.33 | 1.247219 |
| 21. It was easy to make mistakes (Scale Inverted) | 2 | 4 | 2 | 2 | 3 | 3 | 2 | 3.33 | 0.745356 |
| 22. It was easy to recover from mistakes | 5 | 4 | 5 | 4 | 5 | 4 | 1 | 4.50 | 0.5 |
| 23. My mistakes had lasting consequences (Scale Inverted) | 2 | 2 | 1 | 2 | 1 | 1 | 1 | 4.50 | 0.5 |

# APPENDIX D

## Installation

Download minimUML.zip.

The unzipped file should create a directory UML with 3 files:

- ? minimUML.jar – main program file
- ? jdom.jar – library file
- ? umldraw.cfg – configuration file (will be created with defaults if deleted)

Both jar files must be in the same directory.

## Execution

If Java is associated with jar files: Double click minimUML.jar.

From the command line: [Your Java Path]\java.exe –jar minimUML.jar.